# POWER EFFICIENT CARRY PROPAGATE ADDER

Laxmi Kumre[1], Ajay Somkuwar[2] and Ganga Agnihotri[3]

[1,2]Department of Electronics Engineering, MANIT, Bhopal, INDIA
`laxmikumre99@rediffmail.com`
`asomkuwar@gmail.com`
[3]Department of Electronics Engineering, MANIT, Bhopal, INDIA
`gangag@yahoo.com`

## ABSTRACT

*Here we describe the design details and performance of proposed Carry Propagate Adder based on GDI technique. GDI technique is power efficient technique for designing digital circuit that consumes less power as compare to most commonly used CMOS technique. GDI also has an advantage of minimum propagation delay, minimum area required and less complexity for designing any digital circuit. We designed Carry Propagate Adder using GDI technique and compared its performance with CMOS technique in terms of area, delay and power dissipation. Circuit designed using CADENCE EDA tool and simulated using SPECTRE VIRTUOSO tool at 0.18m technology. Comparative performance result shows that Carry Propagate Adder using GDI technique dissipated 55.6% less power as compare to Carry Propagate Adder using CMOS technique.*

## 1. INTRODUCTION

Addition is the most important basic function of any digital processing system. Adders are not only used for arithmetic operation but also necessary to compute virtual physical address in memory fetch operation in all modern computers. Also the adders occupies critical path in key areas of microprocessor, fast adders are prime requirement for the design of fast processing digital system. Many fast adders are available but the design of high speed with low power and less area adders are still challenging. In modern super computers, multiple ALU'S with wide adders and multiple execution core units on the same chip creates thermal hotspots and large temperature gradients. This affects the circuit reliability and increasing the cooling cost of the system. Ideally, adders should have highest performance with least amount of power dissipation and small layout area to minimize unnecessary delays.

With the popularity of portable systems as well as fast growth of power density in integrated circuits, power dissipation becomes main design objectives equal to high performance of the system. For the VLSI designers, designing power efficient adders for digital system has become main goal. Generally Ripple Carry Adders are used among all types of adders because of its compact design but it is the slowest adder. On the other hand, Carry Propagate adders are the fastest adders but they occupy large area and large power dissipation [18].

CMOS is the most common circuit design style/technique for designing any digital circuit but it dissipates most of the power during transistor switching activity. Here we propose a power efficient Carry propagate adder based on gate diffusion input circuit design style. Using this design style, power dissipation in Carry propagate adder is reduced by 55.6% less as compare to CMOS design style. Also it reduces area and propagation delay.

In this paper, next section explains the development of carry propagate adder from ripple carry adder and its design using CMOS technique, Section III explains proposed carry propagate adder based on Gate diffusion input technique. After that section IV gives the details of circuit design simulation using CADENCE EDA tool and then the comparative result with conclusion is explained in section V.

## 2. DESIGN METHODOLOGY OF CARRY PROPAGATE ADDER (CPA)

A Ripple Carry Adder (RCA) is an optimized area-efficient adder design [1]. The layout of a ripple-carry adder is simple, which allows for fast design time but, it is relatively slow, since each full adder must wait for the carry bit to be calculated from the previous full adder. The maximum delay in RCA is computed from the carry-in input to the carry out, passing through each full adder along the way. By making tradeoffs between area and performance delay in adder circuit, faster but larger designs than RCA, can be construct that is Carry Propagate Adder (CPA) [19].

### 2.1. Development of CPA

The computation time of carry in ripple carry adder can be reduced at the price of more complex hardware design of Carry Propagate Adder. The Carry Propagate Adder design can be obtained by creating two main signals P and G for each bit position, depends on whether a carry is propagated through from a less significant bit position, generated in that bit position, or killed in that bit position. In most cases, P is simply the sum output of a half-adder and G is the carry output of the same adder. The carries for every bit position are created after generating P and G signals. Figure 1 shows two different organization of same full adder.

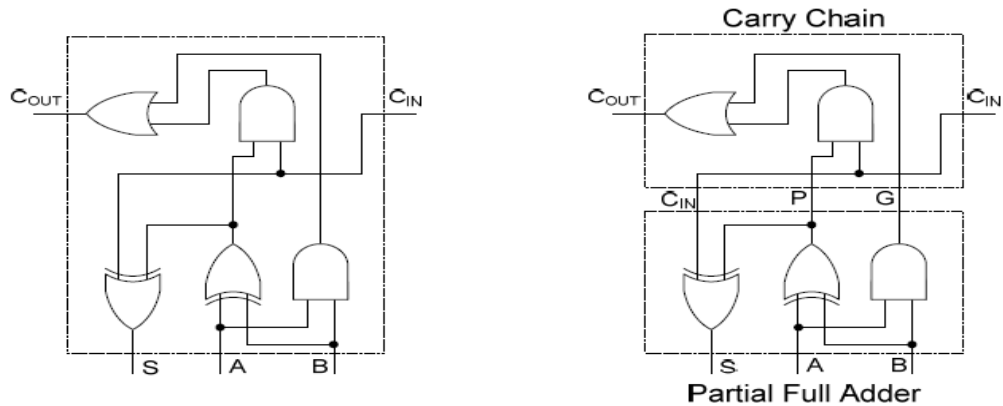

Figure 1: Full adder (left) and Full adder with carry propagates and generates signal (right)

The carry propagate adder design can be obtained by a transformation of the ripple carry design in which the carry logic over fixed groups of bits of the adder is reduced to two level logic. The transformation from ripple carry adder is shown for a 4-bit Carry Propagate adder in figure 2. There are two output signals $P_i$ and $G_i$ from partial full adder to the carry path and one input $C_{IN}$, the carry input from the carry path to each partial full adder. The signal $P_i = A_i$ XOR $B_i$ is called the propagate signal. Whenever $P_i$ is equal to 1, an incoming carry is propagated through the bit position from $C_i$ to $C_{i+1}$. For $P_i$ is equal to 0, carry propagation through the bit is blocked. The function $G_i = A_i$ AND $B_i$ and is called the generate signal. Whenever $G_i$ is equal to 1, the carry output from the position is 1 regardless of the value of $P_i$ and so a carry has been generated in the position. When signal $G_i$ is 0, a carry is not generated, so $C_{i+1}$ is 0 if the carry propagate through the position from $C_i$ is also 0. The propagate and generate signals corresponds exactly to the half adder and they are essential in controlling the values in the ripple carry path. Also as in the full adder, the partial full adder generates the sum function by the XOR of the incoming carry $C_i$ and the propagate signal $P_i$.

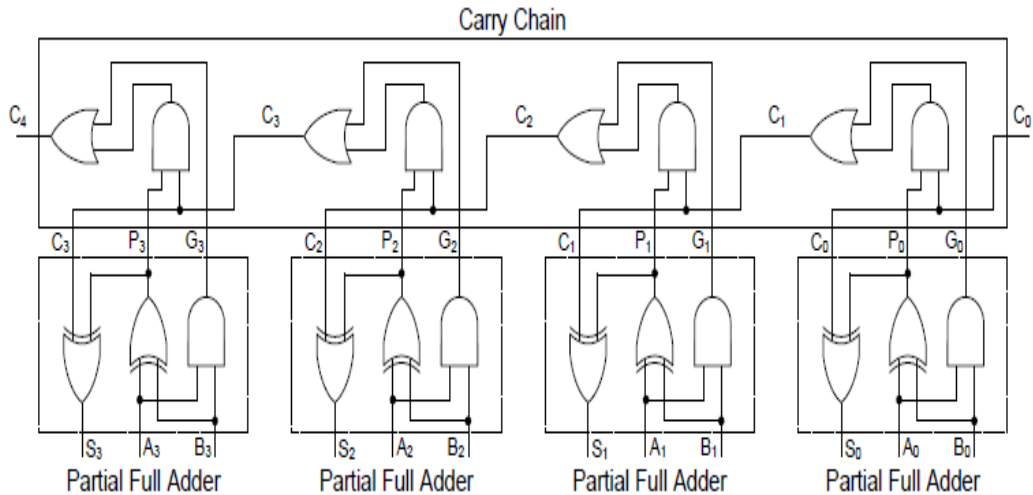

Figure 2: Development of Carry Propagate Adder

The carry chain logic is "multi-level". The Optimized multi-level logic generally results in a smaller but slower circuit than an optimized two level implementation. For a Carry Propagate Adder, convert the multi-level carry chain into a two-level carry chain as shown in figure 3.

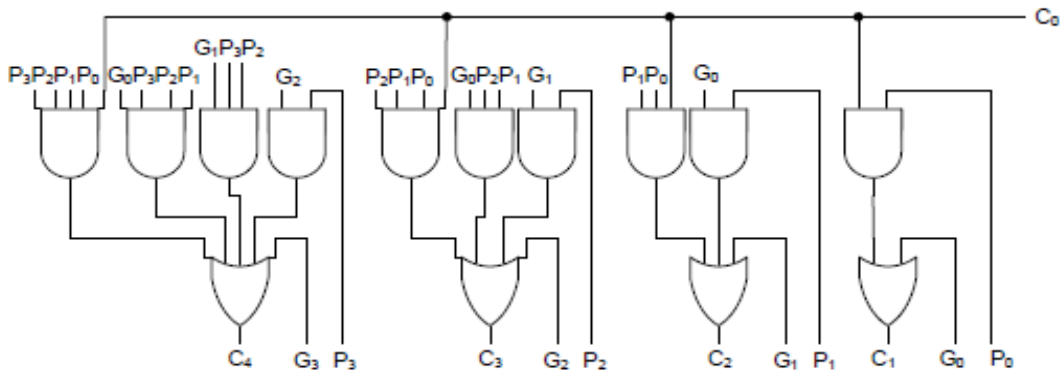

Figure 3: A carry chain block of 4-bit Carry Propagate Adder

Since the logic generating $C_1$ is already two-level, it remains unchanged. The logic for $C_2$ has four levels. So to find the carry propagate logic for $C_2$ is calculated from figure 2 and apply distributive law to obtain

$$C_2 = G_1 + P_1 (G_0 + P_0C_0)$$
$$= G_1 + P_1G_0 + P_1P_0C_0$$

Similarly, the two level logic for $C_3$ can be obtain as

$$C_3 = G_2 + P_2 (G_1 + P_1 (G_0 + P_0C_0))$$
$$= G_2 + P_2G_1 + P_2P_1G_0 + P_2P_1P_0C_0$$

Similarly, the two level logic for $C_4$ can be obtain as

$$C_4 = G_3 + P_3 (G_2 + P_2 (G_1 + P_1 (G_0 + P_0C_0)))$$
$$= G_3 + P_3G_2 + P_3P_2G_1 + P_3P_2P_1G_0 + P_3P_2P_1P_0C_0$$

The carry path remaining in the 4 bit ripple carry adder has a total of eight gates in cascade and so the circuit has a delay of eight gate delays. Since only AND and OR gates are involved in the carry path , the delay from C0 to each of the four carry signals produced C1 through C4, would be just two gate delays.    A 4 bit adder with carry propagate block [18] is shown in figure 4.

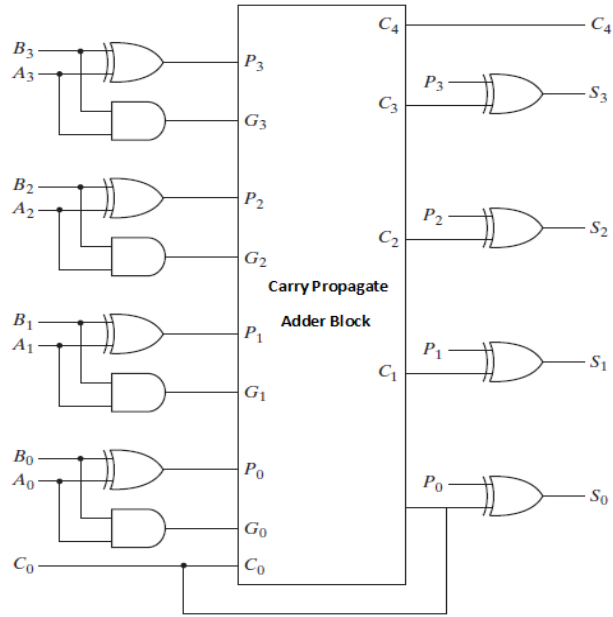

Figure 4:  4 bit Carry Propagate Adder

## 2.2. Design of CPA using CMOS technique

As per the design methodology for the development carry propagate adder, it requires AND, OR and XOR gates to generate sum and carry signal for any binary addition. There are several circuit design techniques at the gate level such as CMOS, Pass Transistor Logic, Domino Logic, ratio based logic etc. Among all gate / transistor level design techniques CMOS is the standard one because of its high noise immunity and low static power consumption. Transistor level design for AND, OR and XOR gates using CMOS techniques [13] are shown in figure 5.

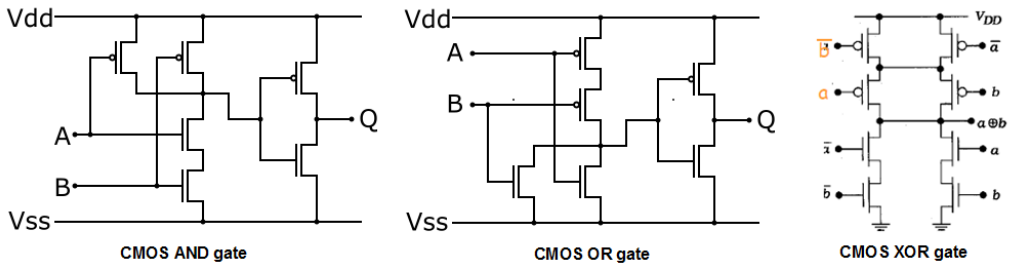

Figure 5: Basic digital gates using CMOS techniques

Carry Propagate Adder can be design using AND, OR and XOR gate of CMOS techniques, but for large digital circuit design CMOS technique is not best at low power VLSI design. Static

CMOS gates are very power efficient because they dissipate almost zero power in idle state. Earlier for designing integrated chips, the power dissipation was not major concern in CMOS devices as the speed and area were dominated design parameters. But as the technology scaling down below the sub-micron levels, the power dissipation per unit area of the chip has become serious issue. The demand of portable battery operated devices also forces to design low power VLSI designs. Basically power consumption in CMOS occurs due to two main components: static dissipation and dynamic dissipation [5] . Static dissipation occurs due to sub threshold conduction when the transistors are off, due to tunnelling current through gate oxide and also due to leakage current through reverse biased diodes. But the amount of static power dissipation is very less as compare to dynamic power dissipation for the digital circuit design.

The dynamic power dissipation [6] in CMOS circuits occurs due to charging and discharging of load capacitance during switching. In one complete cycle, current flows from supply to load capacitance to charge it and then flows from the charged load capacitance to ground during discharge. Multiply by switching frequency on the load capacitance to get the current used and multiply by the voltage again to get the characteristic switching power dissipated by a CMOS device is

$$P = \alpha\, C\, V^2 f$$

Where α is the switching activity factor. Since most gates do not switch at every clock cycle, they are often accompanied by a switching activity factor. Hence the dynamic power consumption can be reduced by reducing the switching activity of any gate. To design low power digital circuits, the new Gate Diffusion Input technique is introduced by A---- in 2001 [ ]. This technique is based on Shannon's expansion theorem and has an advantage of designing any gate using two transistors only. This results less switching activity in any digital operation and consumes less power as compare to CMOS technique.

## 3. CARRY PROPAGATE ADDER BASED ON GDI TECHNIQUE

The GDI technique is first proposed by A. Morgenshtein, A. Fish, and I. A. Wagner in 2001 [2], is based on the use of a simple cell as shown in figure 6. In GDI cell, inputs are applied at source/drain of nMOS and pMOS as well as gate input. There are total three inputs (N, P, and G) with one output. Various logic functions can be performed by using different input combinations at N, P, and G. Table drawn with figure 6 explains all functions.

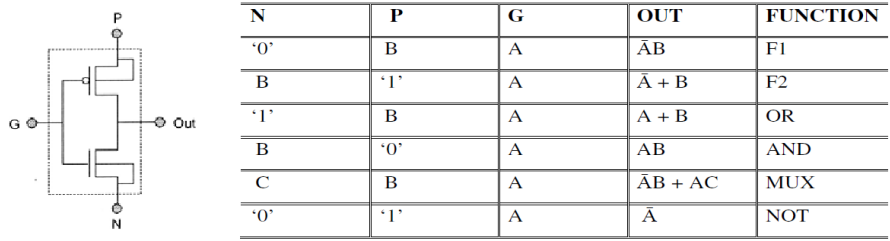

| N | P | G | OUT | FUNCTION |
|---|---|---|-----|----------|
| '0' | B | A | $\bar{A}B$ | F1 |
| B | '1' | A | $\bar{A} + B$ | F2 |
| '1' | B | A | $A + B$ | OR |
| B | '0' | A | $AB$ | AND |
| C | B | A | $\bar{A}B + AC$ | MUX |
| '0' | '1' | A | $\bar{A}$ | NOT |

Figure 6: Basic GDI cell and its various logic functions

The GDI functions given in above table is nothing but simply the extension of a single input CMOS inverter structure into a triple input GDI cell in order to achieve implementation of complicated logic functions with a minimal number of transistors. Extension of any n-input CMOS structure to an (n+ 2) input GDI cell can be done by using P as input instead of supply voltage in the pMOS block of a CMOS structure and an N input instead of ground in the nMOS block. This extended implementation can be represented by the following logic expression [3]:

$$Out = \overline{F(x_1\ldots\ldots x_n)}\, P + F(x_1\ldots\ldots x_n)\, N$$

International Journal of VLSI design & Communication Systems (VLSICS) Vol.4, No.3, June 2013

where $F(x_1......x_n)$ is a logic function of an nMOS block not of the whole original n-input CMOS structure. The above equation is based on Shannon expansion, where any function $F$ can be written as follows:

$$F(x_1......x_n) = x_1 H(x_2......x_n) + \bar{x_1} G(x_2......x_n) = x_1 F(1, x_2......x_n) + \bar{x_1} F(0, x_2......x_n)$$

The output functions of basic GDI cell shown in table are based on Shannon expansion where A, B and C are inputs to G, P and N respectively as,

$$Out = AC + \bar{A}B$$

This fact makes a standard GDI cell very suitable for implementation of any logic function that was written by Shannon expansion. Shannon expansion is a very useful technique for pre-computation based low-power design in sequential logic circuits, due to its multiplexing properties [14]. Hence, GDI cells can be successfully used for low-power design of combinatorial circuits, while combining two approaches - Shannon expansion and combinational logic pre-computation, where transitions of logic input values are prevented from propagating through the circuit if the final result does not change as a result of those transitions.

Using GDI function given in above table, AND, OR and XOR gates are design using only two transistors as shown in figure 7. GDI AND and OR gates require only two transistors [3] whereas CMOS AND and OR gates require six transistors. Also GDI XOR gate uses only four transistors as compare to 12 transistors in CMOS technique. Since the number of transistors required in GDI technique is less, then for any operation of digital gate switching activity will be less and so the power dissipation due to charging and discharging of load capacitance will also be less.

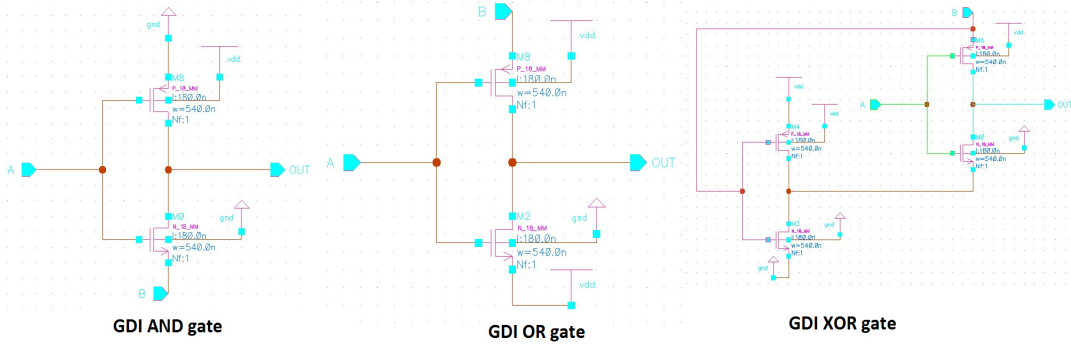

GDI AND gate     GDI OR gate     GDI XOR gate

Figure 7: GDI based digital gates

Table I shows the comparative analysis for AND, OR and XOR gates for both CMOS and GDI technique.

Table 1: Comparative Analysis of GDI and CMOS based digital gates

| Gate type | GDI | | | | CMOS | | | |
|---|---|---|---|---|---|---|---|---|
| | Power (µW) | Delay (nsec) | Power Delay Product | No. of Xsistor | Power (µW) | Delay (nsec) | Power Delay Product | No. of Xsistor |
| AND | 0.149 | 4.948 | 0.737 | 2 | 1.459 | 4.95 | 7.222 | 6 |
| OR | 0.123 | 0.1103 | 0.0135 | 2 | 2.307 | 0.1202 | 0.2773 | 6 |
| XOR | 0.9352 | 0.0161 | 0.0150 | 4 | 1.671 | 0.02291 | 0.0382 | 12 |

From the comparative analysis table it has been observed that as the number of transistors required in GDI based digital gates are less, switching activities are also less and so the total average power dissipation in GDI technique are less as compare to CMOS technique. Also the power – delay product which is considered as a figure of merit correlated with the energy efficiency of a logic gate is the product of power consumption times the duration of the switching event is very – very less in GDI.

## 4. SIMULATION RESULT

Simulation of Carry Propagate Adder using CMOS and GDI are done in CADENCE EDA tool at 180nm technology. All the parameters are set at the time of simulation. W/L ratio is taken as 540/180 nm for the better power delay performance. Simulation is done for Ripple carry adder (RCA) as well as for carry propagate adder (CPA) using both design technique. From the table it is observed that GDI based RCA adder reduces power consumption by 46.75% and CPA by 55.65% as compare to CMOS based adders. Also propagation delay of RCA adder is reduced from 87.02nsec to 65.02nsec and CPDA adder is reduced from 3.118nsec to 3.010nsec by using GDI technique.

Table 2: Comparison of adder design using GDI and CMOS

| Adder Design | GDI | | | | CMOS | | | |
|---|---|---|---|---|---|---|---|---|
| | Power (µW) | Delay (nsec) | Area (µm$^2$) | PDP (FJ) | Power (µW) | Delay (nsec) | Area (µm$^2$) | PDP (FJ) |
| RCA | 46.33 | 65.02 | 5.4432 | 3012.37 | 99.22 | 87.02 | 8.1648 | 8634.124 |
| CPA | 46.25 | 3.010 | 9.72 | 139.21 | 104.3 | 3.118 | 29.16 | 325.207 |

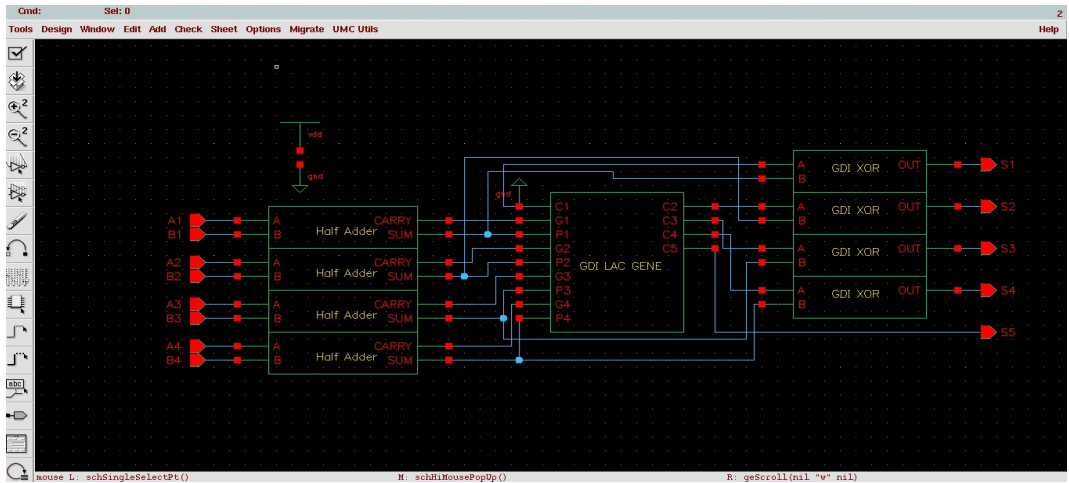

Figure 8: Implementation Carry Propagate Adder based on GDI technique in CADENCE

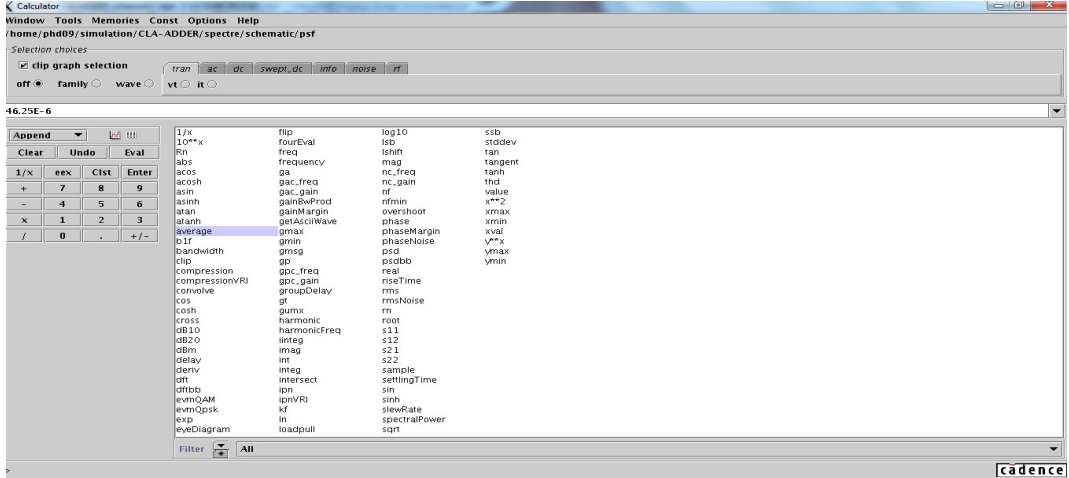

Figure 9: Total average power consumption in Carry Propagate Adder based on GDI technique

## 5. CONCLUSION

In this paper we propose Carry Propagate Adder based on GDI technique. Ripple carry adder and carry propagate adder are simulated using gate diffusion input technique and their performances are compared with CMOS based adders. From the comparative graph it is concluded that the performance of carry propagate adder based on gate diffusion input technique has a better performance than CMOS based adder in terms of power consumption, propagation delay , area required and power delay performance.

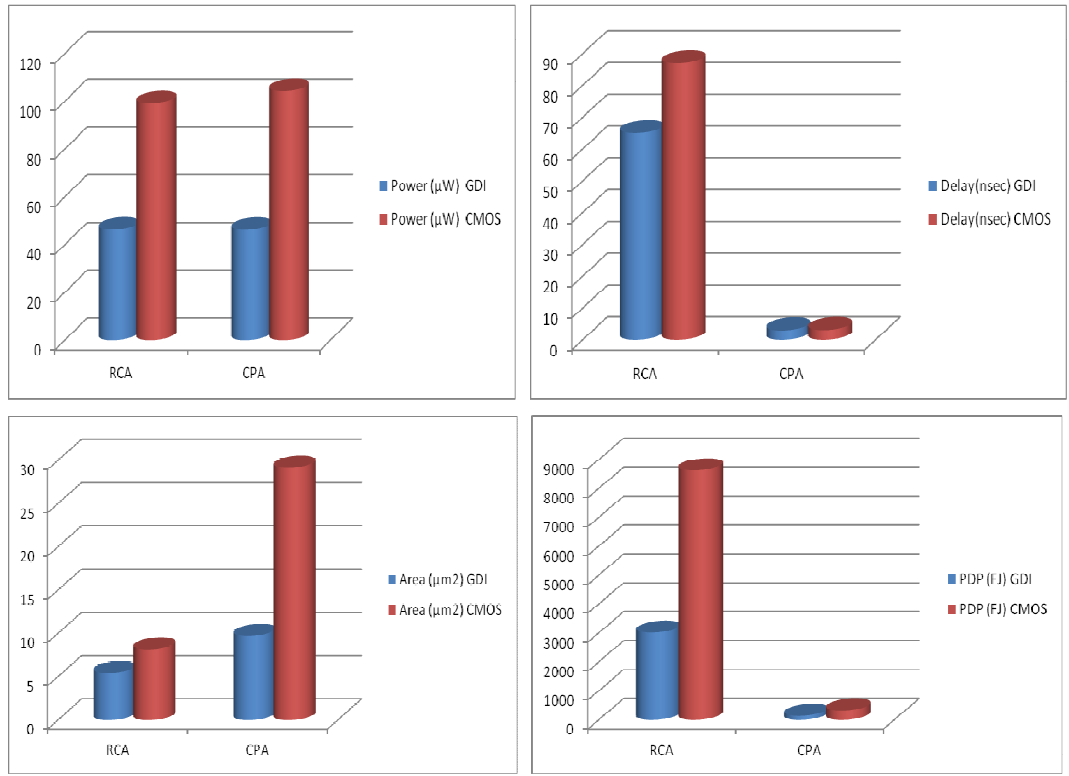

International Journal of VLSI design & Communication Systems (VLSICS) Vol.4, No.3, June 2013

[19] Chetana Nagendra, Mary Jane Irwin and Robert Michael Owens, "Area – Time –Power Tradeoffs in Parallel Adders", IEEE Transactions on circuits and systems- II: Analog and Digital Signal Processing, vol.43, no. 10, October 1996

## Authors

Laxmi Kumre received her B.Tech degree in Electronics and Telecommunication Engineering in 1998, M.Tech. degree in Digital Communication in 2010 and currently pursuing Ph.D degree in Low Power Digital System Design. She is working as Senior Assistant Professor in Department of Electronics and Communication Engineering in MANIT, Bhopal. Her fields of interest are low power design techniques, VLSI Digital system design and Communication Systems. She is fellow member of IEEE, INDIA.

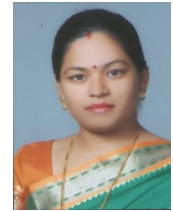

Dr Ajay Somkuwar received his B.Tech degree in Electronics and Telecomm. Engineering , M.Tech. degree in Digital Communication and Ph.D from IIT Delhi. He is working as Professor in Department of Electronics and Communication Engineering in MANIT, Bhopal. He has more than 100 publications and has 22 years experience in teaching and research. He is a member of MIETE and IAENG Professional bodies in INDIA.

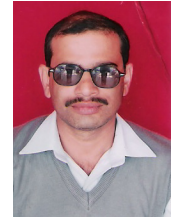

Dr Ganga Agnihotri received her B.Tech degree in Electrical Engineering in 1972, M.Tech. degree in Power Syatem in 1974 and Ph.D in 1989. She is working as Senior Professor in Department of Electrical Engineering in MANIT, Bhopal. Her fields of interest are power system Planning , operation and control. She has more than 100 publications and has 37 years experience in teaching and research. She has a life membership in many Professional bodies in INDIA.

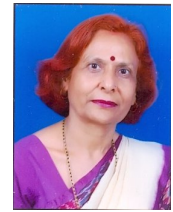

.